\documentclass{appolb}%
\usepackage{amssymb}
\usepackage[usenames,dvipsnames]{color}
\usepackage{graphicx}
\usepackage{makecell}
\usepackage{bm}
\usepackage{amsmath}%
\usepackage{amsfonts}

\newcommand{\eq}[1]{\begin{align} #1 \end{align}}

\definecolor{darkblue}{RGB}{0,0,196}

\begin{document}

\title{Correlations and fluctuations of pions at the LHC
\thanks{Talk presented at XI Workshop on Particle Correlations and Femtoscopy, 3-7 November 2015, Warsaw, Poland.}
}
\author{Viktor Begun$^{1}$
\address{$^{1}$Institute of Physics, Jan Kochanowski University, PL-25406,
Kielce, Poland } }
\maketitle

\begin{abstract}
The intriguing possibility of Bose-Einstein condensation of pions
at the LHC is examined with the use of higher order moments of the
multiplicity distribution. The scaled variance, skewness and
kurtosis are calculated for the pion system. The obtained results
show that the normalized kurtosis has a significant increase for
the case of the pion condensation.
\end{abstract}

\section{Introduction}
The LHC data on mean particle multiplicities~\cite{ALICE} demand
that the temperature at the freeze-out is surprisingly
small~\cite{Petran:2013lja,Stachel:2013zma}. The best fit in the
standard hadron gas model still gives three sigma deviation for
protons~\cite{Stachel:2013zma}. The experimentally measured pion
spectra at the LHC are about 25-50$\%$ higher at low transverse
momenta, $p_T<200$~MeV, than the prediction of the existing
models~\cite{ALICE,hydro}, which worked very well at RHIC. There
are several ways to explain these
deviations~\cite{Petran:2013lja,LHC-expl,Begun-Rybcz}, which one
may group to the equilibrium and the non-equilibrium scenarios,
see~\cite{Begun:2016cva}.
The chemical non-equilibrium allows to reproduce not only the mean
multiplicities, but also the spectra of pions, protons and light
strange particles, including short-living $K^*$ and long-living
$\phi$ mesons~\cite{Begun-Rybcz}.
The source of the non-equilibrium could be the overcooling of the
fireball~\cite{overcooling}, or gluon and then pion condensation
at the LHC~\cite{gluon-cond}. ALICE collaboration reports a large
coherent emission of pions from multi-pion correlation
studies~\cite{ALICE-corr}.
The non-equilibrium parameters obtained in our analysis require
that about $5\%$ of pions are in the Bose-Einstein condensate
(BEC)~\cite{Begun:2015ifa}. The spectra are not sensitive enough
to judge about the existence of BEC, see Fig.~\ref{Fig:1}, left,
from the Ref.~\cite{Begun:2015ifa}. However, multiplicity
fluctuations are infinite at BEC in an infinite
system~\cite{Begun:2005ah}. The fireball, that is created in Pb+Pb
at the LHC, may be large enough to produce the fluctuations that
can be detected.

\section{Pion fluctuations}
\begin{figure}
\centerline{%
 \includegraphics[width=0.48\textwidth]{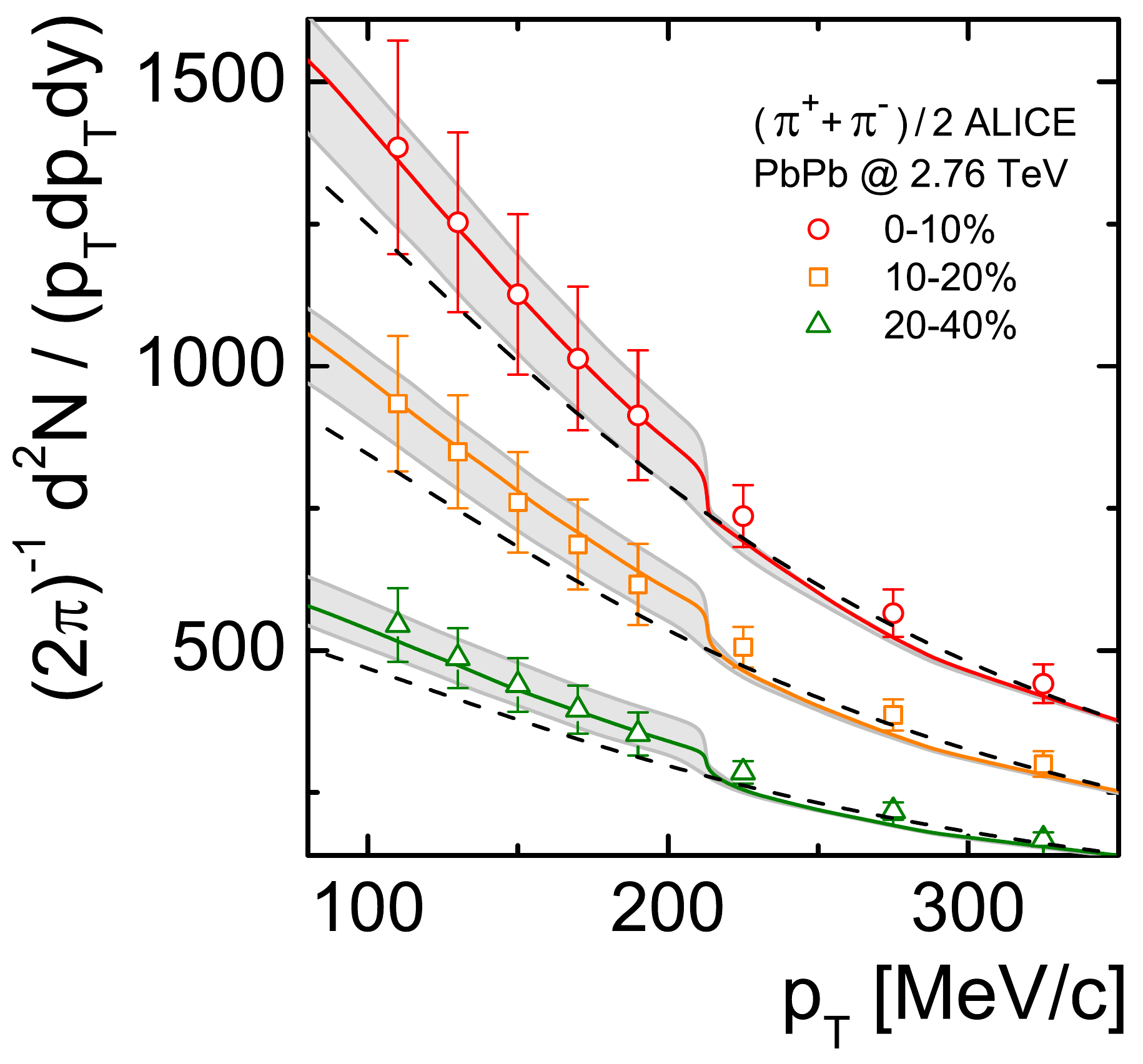}
 \includegraphics[width=0.48\textwidth]{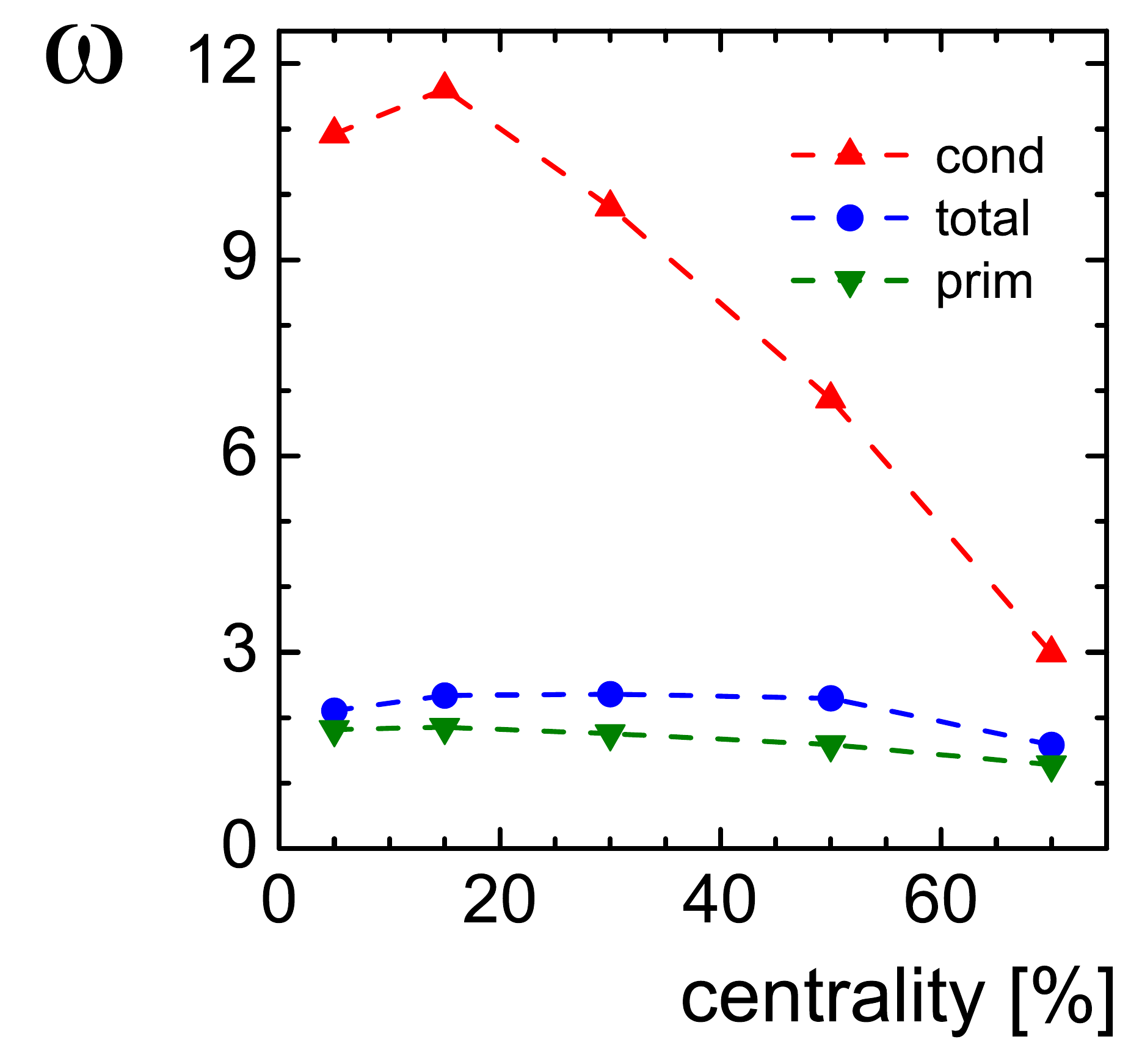}
 }
\caption{Left: The $p_T$ spectra of charged pions in Pb+Pb
collisions with different centralities at the LHC. The dots
correspond to the data, the dashed line - to the equilibrium,
solid line - non-equilibrium with the condensate, the grey area -
to the $10\%$ deviation from the best fit,
see~\cite{Begun:2015ifa} for details. Right: Scaled variance for
positively or negatively charged pions as the function of
centrality for the particles only from the condensate with $p=0$,
cond, for non-condensate primary particles $p\neq0$, prim, and for
the total number of primary particles, $p\geq0$, total.}
\label{Fig:1}
\end{figure}

\begin{figure}
\centerline{%
 \includegraphics[width=0.48\textwidth]{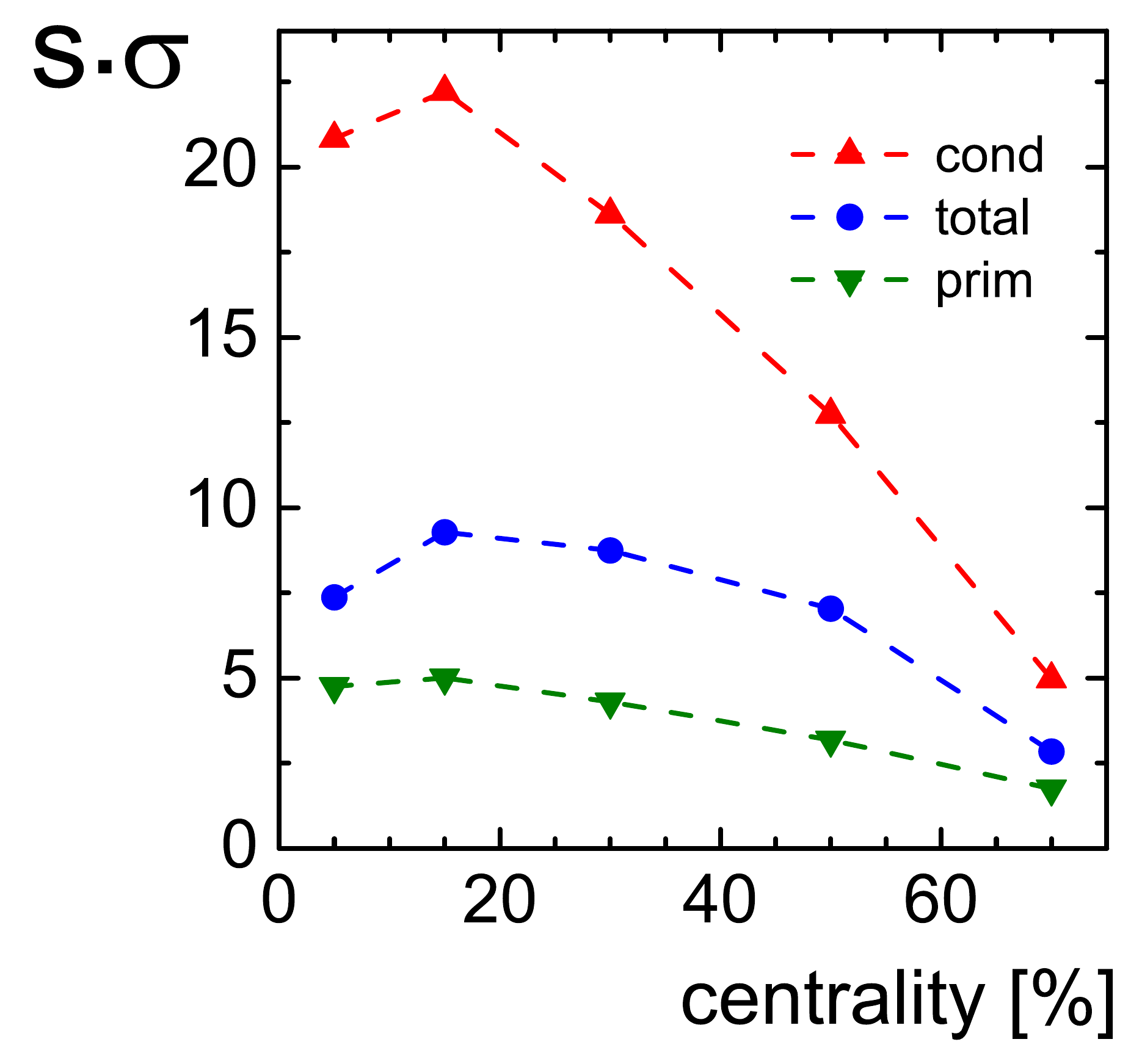}
 \includegraphics[width=0.48\textwidth]{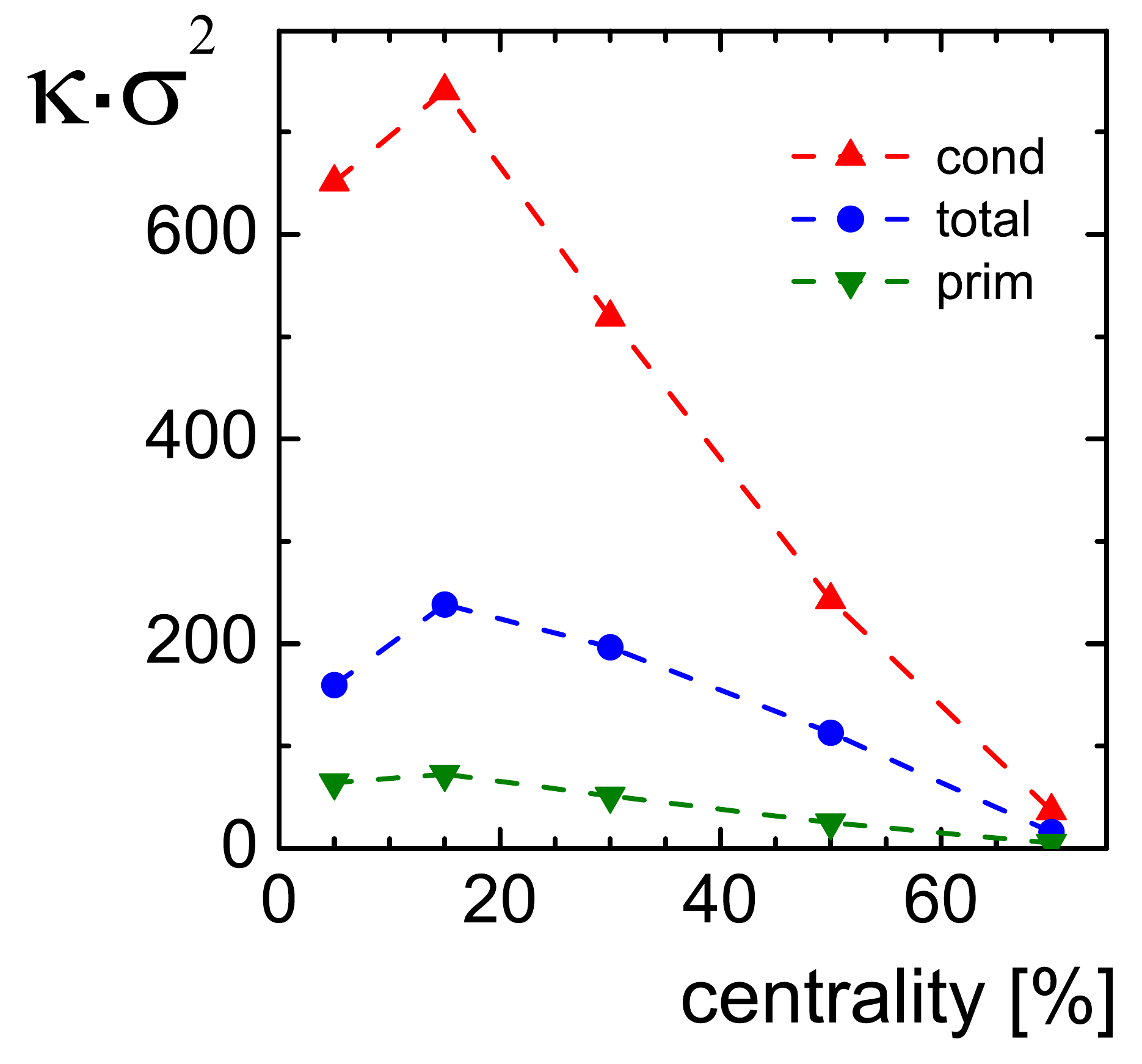}
 }
\caption{The same as Fig.~\ref{Fig:1}, right, for normalized
skewness (left) and normalized kurtosis (right).} \label{Fig:2}
\end{figure}

The fluctuations of primary pions can be calculated analytically,
using the thermodynamic parameters obtained
in~\cite{Begun:2015ifa,Begun:2014aha}. Fluctuations of any order
$\langle(\Delta N)^k\rangle=\langle(N-\langle N\rangle)^k\rangle$
can be expressed as the sum over the momentum levels $\sum_p$ of
the fluctuations on the individual levels $\langle n_p\rangle$
 \eq{
 \langle(\Delta N)^k\rangle ~=~ \sum_p\left(c_1\langle n_p\rangle^1+c_2\langle n_p\rangle^2+\ldots+c_k\langle
 n_p\rangle^k\right)~,
 }
where the coefficients $c_i$ can be straightforwardly calculated
for any order $k$ for primary pions~\cite{Begun:2016cva}. The sum
$\sum_p$ can be replaced by the integral
$(2\pi^2)^{-1}V\int_0^{\infty}p^2dp$ in the infinite volume limit,
$V\rightarrow\infty$, only if there is no BEC. Otherwise, at least
the condensate level should be taken into
account~\cite{Begun:2008hq}.
As a thought experiment, one can separately calculate fluctuations
on the $p=0$ level, on $p\neq0$ levels, and the fluctuations from
all $p\geq0$ levels. These cases are correspondingly labelled as
cond, prim, and total in Figs.~\ref{Fig:1}-\ref{Fig:2}.
One can see, that the signal can be very large\footnote{For random
emission of particles one has Poisson distribution, for which
$\omega=S\cdot\sigma=\kappa\cdot\sigma^2=1$. For large number of
particles Poisson is often replaced by Gauss distribution, which
similarly has $\omega=1$, but
$S\cdot\sigma=\kappa\cdot\sigma^2=0$.}, and it is larger for
higher order fluctuations. The fluctuations on the $p=0$ level are
the order of magnitude larger than the fluctuations on the
$p\neq0$ levels. The total fluctuations of primary pions are in
between, closer to the $p\neq0$ case.

\section{Conclusions}
In order to increase the possible signal of BEC, one may try to
find a way to select more particles from the condensate. The
easiest way to do it, is to impose a momentum cut, because pions
from the condensate receive a finite momentum boost due to
movement with the hypersurface\footnote{The maximal allowed
momentum for particles from the condensate is shown as the step in
Fig.~\ref{Fig:1}, right. The step itself is the artefact of the
approximation that we have no excited levels above the condensate.
An inclusion of a few of them would produce a few steps and a
smoother line.}~\cite{Begun:2015ifa}. A realistic estimate should
also include the contribution from resonance decays,
see~\cite{Begun:2016cva}.

\bigskip

\textbf{Acknowledgments:} I thank W.~Florkowski, M.I.~Gorenstein,
and L.~Tinti for useful discussions. This research was supported
by Polish National Science Center grant
No.~DEC-2012/06/A/ST2/00390.

\end{document}